\documentclass[]{iopart}

\usepackage{graphicx}
\usepackage{multirow}
\usepackage{rotating}
\usepackage{iopams}

\newcommand{\sci}[2]{\ensuremath{#1\times 10^{#2}}}
\newcommand{\sub}[1]{\ensuremath{_{\rm #1}}} 
\newcommand{\super}[1]{\ensuremath{^{\rm #1}}} 
\newcommand{\url}[1]{{\tt #1}}
\renewcommand{\vec}[1]{\boldsymbol{#1}}
\renewcommand{\hat}[1]{\boldsymbol{#1}}

\begin{document}

\title[Nonlinear fluid response in JDFT studies of battery systems]
{The importance of nonlinear fluid response in\\joint density-functional theory studies\\of battery systems}

\author{Deniz Gunceler, Kendra Letchworth-Weaver, Ravishankar Sundararaman, Kathleen A Schwarz and T A Arias}
\address{Cornell University Department of Physics, Ithaca, NY 14853, USA}
\date{\today}



\begin{abstract}
Delivering the full benefits of first principles calculations to battery materials 
demands the development of accurate and computationally-efficient electronic structure methods
that incorporate the effects of the electrolyte environment and electrode potential.  
Realistic electrochemical interfaces containing polar surfaces are beyond
the regime of validity of existing continuum solvation theories developed
for molecules, due to the presence of significantly stronger electric fields.
We present an \emph{ab initio} theory of the nonlinear dielectric and ionic
response of solvent environments within the framework of joint density-functional theory,
with precisely the same optimizable parameters as conventional polarizable continuum models.
We demonstrate that the resulting nonlinear theory agrees with the standard linear
models for organic molecules and metallic surfaces under typical operating conditions.
However, we find that the saturation effects in the rotational response of polar solvent
molecules, inherent to our nonlinear theory, are crucial for a qualitatively correct
description of the ionic surfaces typical of the solid electrolyte interface.
\end{abstract}
\maketitle

\section{Introduction} \label{sec:Introduction}
The development of better batteries is a critical step towards
reducing energy reliance from oil to renewable energy resources.  
Experimental studies have historically
made great progress in identifying promising battery materials
\cite{Battery-Review}, but a number of challenges
remain.  For instance, the solid electrolyte interfaces (SEI) \cite{SEI} of the
electrodes play a crucial role in the thermodynamics and kinetics of
battery operation, but these surfaces are currently poorly
understood.  Experiments have yet to even conclusively
determine the compositions of many of these surfaces \cite{SEI-review}.  
Other important challenges include understanding reaction mechanisms at the
surfaces of electrodes and identifying new electrode materials.
These problems are well-suited for computational study, which can complement an
experimental approach through inexpensive, rapid but accurate calculations.  

However, electrochemical systems pose a unique challenge for theoretical studies:
the processes of interest occur at an interface that requires
simultaneous quantum-mechanical and statistical treatment.
The electrode and reactants on its surface need to be described
with a quantum-mechanical method in order to
capture the level of detail required to predict chemical reactions.
The liquid electrolyte plays an equally important role in determining the
reaction pathways, and necessitates a statistical treatment
due to the need to sample the configuration space of the liquid.

The most straightforward approach to a combined statistical and
quantum mechanical calculation is \emph{ab initio} molecular
dynamics \cite{CPMD}, which is expensive since adequate
statistical sampling necessitates several thousands of steps
at the electronic structure level of detail.
This cost may be ameliorated by combining classical molecular dynamics
with electronic structure only for relevant parts of the system, as in
the Quantum Mechanics / Molecular Mechanics (QM/MM) methods \cite{QMMM}.
However, statistical sampling issues and the need for
coupling constant integration for estimating free energies complicate
the analysis of the results of any molecular dynamics based method.

The complexity and computational cost due to statistical sampling can be
avoided by working directly with equilibrium properties of the system.
Joint density-functional theory (JDFT) \cite{JDFT} is an exact
variational principle for the free energy of an electronic system
in contact with a liquid, in terms of the densities of the two subsystems.
This framework enables systematic approximations such as combining
electronic density-functional theory for the system of interest
with classical density-functional theory for the liquid environment.

Polarizable continuum models (PCM's) \cite{PCM-Review} are a class of
highly efficient simplified theories where the effect of the fluid is captured
by placing the electronic system in an appropriately chosen dielectric cavity,
optionally with corrections for physical effects such as cavitation energies
and dispersion interactions. However, the efficiency of these models comes
at the cost of empiricism and a loss of key physical features of the fluid.

The empiricism of PCM approaches has been partially mitigated by
constructing variants of the model \cite{JDFT, PCM-Kendra} that are
highly simplified approximations within the framework of JDFT.
So far, PCM approximations \cite{JDFT,PCM-Kendra,PCM-Gygi,PCM-Marzari}
have replaced the fluid with a \emph{linear} dielectric response
which turns out to be adequate for the solvation of most molecules and
some surfaces, such as those of metals. However, the highly polar 
surfaces typical of battery systems impose strong electric fields
on solvents that invoke a highly nonlinear response;
linear response approximations lead to qualitatively incorrect results
as we demonstrate in Section~\ref{sec:IonicSurfaces}.

In this paper, we present a systematic framework (Section~\ref{sec:Framework})
for developing PCM-like approximations within joint density-functional theory,
and use it to construct a nonlinear polarizable continuum model
(Sections~\ref{sec:NonlinearDielectric} and \ref{sec:NonlinearIons})
that is both inexpensive and sufficiently accurate to account for
complex reactions, including those occurring on ionic surfaces.
We show that the nonlinear dielectric model reproduces molecule solvation energies
(Section~\ref{sec:Molecules}), and with the inclusion of nonlinear ions, 
potentials of zero charge for metallic surfaces (Section~\ref{sec:MetalSurfaces})
with accuracy similar to that of the linear model. 
Finally, we demonstrate that the inclusion of nonlinear dielectric saturation effects
facilitates accurate predictions for ionic surfaces in solution (Section~\ref{sec:IonicSurfaces}),
making this model particularly suited for theoretical studies of battery materials.

\section{Nonlinear polarizable continuum model} \label{sec:Theory}

\subsection{Joint density-functional theory framework for polarizable continuum models} \label{sec:Framework}

The fundamental quantity of interest for \emph{ab initio} studies
of electrochemical and other solvated systems is the free energy of a
quantum mechanical system in \emph{equilibrium} with a liquid environment.
Therefore, the most direct route to this quantity is a theory
in terms of the equilibrium densities of the two subsystems.
Joint density-functional theory (JDFT) \cite{JDFT} is based on
an exact variational principle for this free energy in terms 
of these equilibrium densities, and provides a rigorous framework
for the development of practical approximations.

The total free energy of such a system may be exactly partitioned as
\begin{equation}
A\sub{JDFT}[n,\{N_\alpha\}]
 = \underbrace{A\sub{HK}[n]}\sub{electronic}
 + \underbrace{\Phi\sub{lq}[\{N_\alpha\}]}\sub{liquid}
 + \underbrace{\Delta A[n,\{N_\alpha\}]}\sub{coupling},
\label{eqn:JDFT}
\end{equation}
where $A\sub{HK}$ is the exact Hohenberg-Kohn electronic density
functional \cite{HK-DFT}, $\Phi\sub{lq}$ is the exact free energy
functional of the liquid \cite{ThermalDFT-Mermin}, and the remainder,
$\Delta A$, is the free energy for the interaction of the two systems.
Minimizing the above functional yields the ground state
electron density $n(\vec{r})$ and the set of nuclear densities
$\{N_\alpha(\vec{r})\}$ for the fluid.

In practice, each of the in-principle exact pieces of (\ref{eqn:JDFT}) needs to be approximated,
and the power of the framework lies in the capability of independently
selecting the level of approximation for each piece depending on
the type of system, desired accuracy and available computational resources.
The electronic system may be treated within the Kohn-Sham formalism \cite{KS-DFT}
with any of the standard exchange-correlation functionals, or if necessary,
with correlated quantum chemistry methods or quantum Monte Carlo methods
as demonstrated in \cite{Katie-QMC}.

The liquid free energy may be treated within the rigid molecule
classical density-functional theory formalism \cite{RigidMoleculeCDFT},
with an approximation for the excess free energy of the liquid;
reliable functionals for liquid water have been constructed
from its equation of state \cite{RigidMoleculeCDFT,BondedVoids}
and functionals for other liquids are available in the literature. (See \cite{CDFT-Survey} for a survey.)
The interaction of the two subsystems, $\Delta A$, may be treated using a density-only
electronic density functional approach \cite{DODFT-Coupling}.
These approximations may be independently improved or simplified,
as required for the system of interest.

Using a classical density-functional theory for the liquid within JDFT
is a powerful tool for studying solvated electronic systems.
However, the complexity of the theory can occasionally obscure an intuitive physical
interpretation of the results. This intuition may be better obtained from simpler
and possibly less accurate versions of the theory that capture the bare minimum
of physical effects required to describe the systems and properties of interest.

Polarizable continuum models (PCM) are highly simplified theories that account
for liquid effects by embedding the electronic system in a dielectric cavity.
The linear response approximation in PCM, however, is inadequate for
the study of electrochemical systems that involve liquids in strong electric fields.
Here, we develop a general framework for constructing PCM-like approximations
within joint density-functional theory, which we use in the following sections to
construct a nonlinear PCM with the same optimizable parameters as those of the linear model.

We start by dividing the liquid contributions to the free energy functional
into physical effects assumed to be separable in polarizable continuum models,
and rewrite the last two terms of (\ref{eqn:JDFT}) in the following form \footnote{
Here and throughout this paper, we use atomic units $4\pi\epsilon=e=\hbar$=$m_e$=$k_B$=1.}
\begin{eqnarray}
\fl A\sub{diel} \equiv \Phi\sub{lq} + \Delta A \nonumber\\
\fl\qquad = A_\epsilon[s,\vec{\varepsilon}] + A_\kappa[s,\mu] 
	+ \int \rmd\vec{r} \int \rmd\vec{r}' \frac{\rho\sub{lq}(\vec{r})}{|\vec{r}-\vec{r}'|}
	\left(\rho\sub{el}(\vec{r}) + \frac{\rho\sub{lq}(\vec{r})}{2} \right)
	+ A\sub{cav}[s].
\label{eqn:JDFT-PCM}
\end{eqnarray}
Dielectric response dominates the electrostatic interaction of a fluid consisting of
neutral molecules alone, and the first term $A_\epsilon$ captures the corresponding internal energy.
In addition to neutral solvent molecules, electrolytes typically include charged ions
that contribute an additional monopole response. The optional term, $A_\kappa$,
accounts for the internal energy of the ions if present in the solution.
The densities of the molecules and ions of the solvent are modulated by the cavity shape
function $s(\vec{r})$, which in turn is determined by the electron density $n(\vec{r})$.

The third term of (\ref{eqn:JDFT-PCM}) is the mean field electrostatic interaction
of the liquid bound charge density $\rho\sub{lq}$ with itself and
the electronic system of total charge density $\rho\sub{el}$.
Here, $\rho\sub{lq} \equiv \rho_\epsilon + \rho_\kappa$ includes dielectric
and ionic contributions, while $\rho\sub{el} \equiv n + \rho\sub{nuc}$ includes
contributions from the electrons and the nuclei (or pseudopotential cores)
of the subsystem treated using electronic density-functional theory.
The contributions from all remaining effects of the fluid, such as cavitation and dispersion,
are gathered into the final term of (\ref{eqn:JDFT-PCM}), $A\sub{cav}$,
and are assumed to depend only on the shape of the cavity $s(\vec{r})$.
We detail specific approximations for each of these terms in the following subsections.

So far, the dielectric and ionic responses are still fully general, except for the
mean-field assumption in their interaction with each other and the electronic system.
In reality, these responses are nonlinear as well as nonlocal, while
conventional polarizable continuum models \cite{JDFT, PCM-Review, PCM-Kendra, PCM-Gygi, PCM-Marzari} assume both linearity and locality.
In this work, we retain the local response approximation,
but develop a nonlinear theory for dielectric and ionic response
in sections~\ref{sec:NonlinearDielectric} and \ref{sec:NonlinearIons}.
We obtain a linear PCM comparable to \cite{PCM-Kendra} and \cite{PCM-Marzari}
in Section~\ref{sec:LinearPCM} as the low-field limit of our general nonlinear theory.

\subsection{Cavity shape function $s(\vec{r})$, and dependent energy $A\sub{cav}$} \label{sec:Cavitation}

Polarizable continuum models replace the liquid with a dielectric cavity
surrounding the electronic system. In variants of the model suitable
for treating solid surfaces (which typically require a plane-wave basis),
the dielectric constant is smoothly switched from the vacuum value of 1
to the bulk liquid value $\epsilon_b$ \cite{JDFT,PCM-Kendra,PCM-Gygi,PCM-Marzari}.
The spatial modulation of the dielectric constant may be written as
$\epsilon(\vec{r}) = 1 + (\epsilon_b-1) s(\vec{r})$, so that 
$s(\vec{r}) \in [0,1]$ describes the shape of the cavity.

Further, these variants of PCM assume that the cavity shape $s(\vec{r})
= s(n(\vec{r}))$ is determined entirely by the local electron density.
The exact functional form of $s(n)$ is not important as long as it switches
smoothly between 0 at high electron densities and 1 at low electron densities,
and rapidly approaches the extreme values away from the transition region.
Following \cite{JDFT}, for the rest of this work, we use
\begin{equation}
s(n) = \frac{1}{2}\textrm{erfc} \frac{\log(n/n_c)}{\sigma\sqrt{2}}
\label{eqn:ShapeFunction}
\end{equation}
where the parameter $n_c$ sets the critical electron density around which the cavity
smoothly `switches on', and $\sigma$ controls the width of that transition.

In the following subsections, we develop an \emph{ab initio} theory
for the nonlinear dielectric and ionic response of solvents,
which we find to be the dominant effects at the charged or highly polar surfaces
in electrochemical systems due to the strong electric fields.
The cavity shape function, however, includes unknown parameters
that are typically fit \cite{PCM-Review, PCM-Marzari} to reproduce
the solvation energies of small organic molecules.
These solvation energies are sensitive to other free energy contributions such as
cavitation and dispersion, which although negligible in the electrochemical systems
of interest, cannot be ignored during the determination of fit parameters.

These additional free energy contributions have complicated dependences on the shape of the cavity,
for which several empirical approximations have been developed (see \cite{PCM-Review} for a review).
Andreussi and coworkers \cite{PCM-Marzari} demonstrated that a simple empirical model
expressing the sum of all these effects as an effective surface tension for the cavity
works reasonably well for the solvation energies of small organic molecules.
Since we need the additional effects only as auxiliary contributions during the fit
to the molecular solvation data, we adopt their simplified model here by writing
\begin{equation}
A\sub{cav}[s]
	= \tau \underbrace{\int \rmd\vec{r} \left|\vec{\vec{\nabla}}{s}\right| }_S,
\label{eqn:cavitation}
\end{equation}
where $S$ is a surface area estimate for the cavity described by $s(\vec{r})$,
and $\tau$ is an effective tension that is determined by the fit to solvation energies.

\subsection{Nonlinear dielectric internal energy, $A_\epsilon$}\label{sec:NonlinearDielectric}

The dielectric response of liquids includes contributions from molecular
polarizability as well as rotations of molecules with permanent dipole moments.
The response of highly polar solvents such as water is dominated by rotations.
With increasing field strength, the molecular dipoles increasingly align
with the electric field, eventually saturating the rotational response.
The polarizability response, which includes electronic polarizability and
flexing modes of the molecules, typically becomes stronger at higher fields.
It is therefore important to consider all these contributions even for
solvents whose linear response is dominated by rotations.

The typical electric fields encountered in solvation can significantly
saturate the rotational response of solvents, but are usually insufficient
to access the nonlinear regime of the remaining contributions.
We therefore split the internal energy of the dielectric $A_\epsilon$
into rotational $A\sub{rot}$ and polarization $A\sub{pol}$ parts,
and construct a nonlinear theory for the rotational part alone.

Within the polarizable continuum ansatz, the liquid consists of
molecules distributed with the bulk density $N\sub{mol}$
modulated by the cavity shape function $s(\vec{r})$.
The internal energy corresponding to linear polarization response
with an effective molecular polarizability $\chi\sub{mol}$ is
\begin{equation}
A\sub{pol}[\vec{P}\sub{pol}] = \int \rmd\vec{r} N\sub{mol} s(\vec{r})
	\frac{1}{2} \chi\sub{mol} \vec{P}\sub{pol}^2(\vec{r}),
\label{eqn:Apol}
\end{equation}
where $\vec{P}\sub{pol}(\vec{r})$ is the induced dipole moment per molecule.
This dipole moment contributes a bound charge, $\rho\sub{pol}(\vec{r}) =
-\vec{\nabla}\cdot (N\sub{mol}s(\vec{r})\vec{P}\sub{pol}(\vec{r}))$.

Physically, the nonlinearity of the rotational response arises from a competition
between the rotational entropy of the molecules and their interaction
with the self-consistent electric field. We therefore begin with the exact
rotational entropy for an ideal gas of dipoles with the cavity-prescribed
density $N\sub{mol}s(\vec{r})$ at temperature $T$, then approximate rotational correlations, and write
\begin{equation}
\fl A\sub{rot}[p_{\hat{e}},l] = \int \rmd\vec{r} TN\sub{mol} s(\vec{r})
\left[
	\int \frac{\rmd\hat{e}}{4\pi} p_{\hat{e}} \log p_{\hat{e}}
	- l(\vec{r}) \left( \int \frac{\rmd\hat{e}}{4\pi} p_{\hat{e}} - 1 \right)
	- \frac{\alpha\vec{P}^2\sub{rot}(\vec{r})}{2} 
\right].
\label{eqn:Arot}
\end{equation}
Here, $p_{\hat{e}}(\vec{r})$ is the probability that a molecule
at location $\vec{r}$ has its dipole oriented along unit vector $\hat{e}$,
and the second term of (\ref{eqn:Arot}) constrains the normalization of
$p_{\hat{e}}(\vec{r})$ with Lagrange multiplier field $l(\vec{r})$.

The final term of (\ref{eqn:Arot}) captures the correlations in
dipole rotations within a `local polarization density approximation (LPDA)'.
We choose the simplest possible form for this correction,
quadratic in the local dimensionless polarization $\vec{P}\sub{rot}(\vec{r})
= \int \frac{\rmd\hat{e}}{4\pi} p_{\hat{e}}(\vec{r}) \hat{e}$,
and constrain the prefactor $\alpha$ to reproduce the
bulk linear dielectric constant $\epsilon_b$.
Finally, the rotational response contributes a bound charge
$\rho\sub{rot}(\vec{r}) = -\vec{\nabla}\cdot
(p\sub{mol}N\sub{mol}s(\vec{r})\vec{P}\sub{rot}(\vec{r}))$
within the local response approximation, where $p\sub{mol}$
is the permanent molecular dipole moment.

The Euler-Lagrange equation for minimizing the total free energy with respect to $p_{\hat{e}}$
implies that, at equilibrium, the orientation distribution must be of the form $p_{\hat{e}} =
\exp(\vec{\varepsilon}\cdot\hat{e})$ for some vector field $\vec{\varepsilon}(\vec{r})$.
Using the remaining Euler-Lagrange equations to eliminate $\vec{P}\sub{pol}(\vec{r})$ and $l(\vec{r})$ in
favor of  $\vec{\varepsilon}(\vec{r})$, the sum of (\ref{eqn:Apol}) and (\ref{eqn:Arot}) simplifies to
\begin{equation}
\fl A_\epsilon[\vec{\varepsilon}(\vec{r})] =  \int \rmd\vec{r} TN\sub{mol} s(\vec{r}) \left[
	\varepsilon^2 \left( f(\varepsilon) - \frac{\alpha}{2} f^2(\varepsilon)
	+ \frac{X}{2} (1-\alpha f(\varepsilon))^2 \right) - \log\frac{\textrm{sinh}\varepsilon}{\varepsilon}
\right],
\label{eqn:Aeps}
\end{equation}
with corresponding dielectric bound charge
\begin{equation}
\rho_\epsilon(\vec{r}) = - \vec{\nabla} \cdot \left[
	p\sub{mol} N\sub{mol} s(\vec{r}) \vec{\varepsilon}
	\left( f(\varepsilon) + X (1-\alpha f(\varepsilon)) \right)
\right].
\label{eqn:rhoEps}
\end{equation}
Here, $f(\varepsilon) = (\varepsilon~\textrm{coth}~\varepsilon-1)/\varepsilon^2$
is the effective dimensionless rotational susceptibility defined by $\vec{P}\sub{rot}
= f(\varepsilon) \vec{\varepsilon}$, and $X \equiv \chi\sub{mol}T/p\sub{mol}^2$
is its counterpart for the linear polarization response.  

The resulting theory for the dielectric has four solvent-dependent
parameters ($N\sub{mol}$, $p\sub{mol}$, $X$ and $\alpha$),
of which the bulk molecular density, $N\sub{mol}$, is directly measurable.
The effective molecule dipole moment in the liquid, $p\sub{mol}$, differs from the
gas-phase value, and in principle, can be determined from measurements of
the lowest order nonlinear response coefficient \cite{pMolFromNonlinearEps}.
However, such measurements are difficult and not readily available for most solvents.
Instead, we compute $p\sub{mol}$ as the self-consistent dipole moment of
a single solvent molecule in a solvated \emph{ab initio} calculation
employing a nonlinear polarizable continuum description of the same solvent.
The resulting dipole moment for the solvents used in this work are listed
in table \ref{tab:solvents}. Note that $p\sub{mol}$ is larger than the gas phase dipole
moment for all these solvents because charged centers in the molecule are surrounded
by bound charges of the opposite sign which favor an increase in the polarization,
as shown for water in figure~\ref{fig:waterBoundCharge}. We also find that for water,
$p\sub{mol} = 0.94ea_0$ gratifyingly agrees with the SPC/E molecular dynamics
model value of $0.92ea_0$ \cite{SPCE}, in contrast to the gas phase value of $0.73ea_0$.

The remaining solvent-dependent parameters, the correlation factor for rotations $\alpha$
and the effective dimensionless polarizability $X$, are constrained to reproduce
the bulk static and high frequency dielectric constants, $\epsilon_b$ and
$\epsilon_\infty$ respectively. Using the bulk linear response of the above functional,
we can analytically show that
\begin{equation}
X = \frac{T(\epsilon_\infty-1)}{4\pi N\sub{mol}p\sub{mol}^2} \quad\textrm{and}\quad
\alpha = 3 - \frac{4\pi N\sub{mol}p\sub{mol}^2}{T(\epsilon_b-\epsilon_\infty)},
\label{eqn:epsBulk}
\end{equation}
since the rotational response freezes out and does not contribute to $\epsilon_\infty$.
In principle, $\epsilon_\infty$ should be the dielectric constant at infrared frequencies between the
rotational and vibrational resonances, but in the absence of experimental data in that frequency regime,
we use the readily measurable optical dielectric constant, which is the square of the refractive index.

\begin{figure}
\center{\includegraphics[width=3.4in]{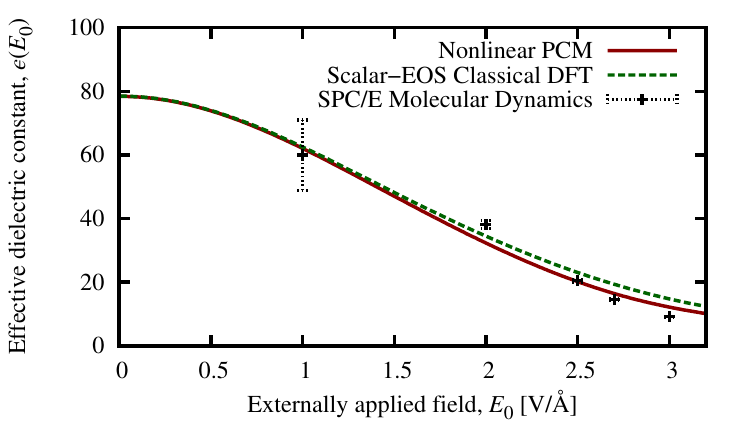}}
\caption{Comparison of the effective dielectric constant of water
as a function of uniform externally applied field $E_0$ for the
nonlinear PCM model (\ref{eqn:Aeps}) compared to
SPC/E molecular dynamics results \cite{NonlinearEpsSPCE} and
classical density functional predictions \cite{RigidMoleculeCDFT}.
The effective dielectric constant is defined by $\epsilon = E_0/E$,
where $E = E_0 - 4\pi P$ is the net electric field including the
screening due to the dielectric polarization density $P$.
Within this theory, this response is determined entirely by
bulk liquid properties $\epsilon_b$, $\epsilon_\infty$ and $N\sub{mol}$,
along with the molecule dipole moment $p\sub{mol}$ obtained from
a self-consistent ab initio calculation solvated with the present model.
\label{fig:NonlinearEps}}
\end{figure}

We have therefore produced a density-functional theory for the nonlinear dielectric response
of an arbitrary solvent constrained entirely by measurable macroscopic properties.
The response at field strengths relevant for solvation is not accessible
experimentally, but it has been estimated using molecular dynamics.
Figure~\ref{fig:NonlinearEps} demonstrates that the bulk nonlinear dielectric response
of the present theory is in excellent agreement with molecular dynamics results
for water\cite{NonlinearEpsSPCE} using the SPC/E pair potential model \cite{SPCE}.
The present theory, which uses LPDA for rotational correlations,
produces essentially the same nonlinear response to uniform electric fields as classical
density functional theories with the scaled mean-field electrostatics approximation \cite{LischnerH2O}. 
The minor differences between the present theory and the classical density
functional results \cite{RigidMoleculeCDFT} shown in Figure~\ref{fig:NonlinearEps}
are due to electrostriction; the latter theory accounts for changes in the
equilibrium fluid density in the presence of a strong uniform electric field.

\subsection{Nonlinear ionic system internal energy, $A_\kappa$} \label{sec:NonlinearIons}

The previous section derived the dielectric response of liquids from the dipolar
rotational and polarization response of liquid molecules to the local electric field.
Ionic species in the liquid introduce Debye screening by contributing an additional
monopolar response, which changes the local ionic density in response to
the local electric potential. A simple description of this response at the linearized
Poisson-Boltzmann level suffices for many electrochemical systems \cite{PCM-Kendra}.
For the electrode-electrolyte interface, this level of theory corresponds roughly
to the Gouy-Chapman-Stern model, but misses the nonlinear capacitance effects due to ion
adsorption. Here, we explore whether a full Poisson-Boltzmann treatment
within the polarizable continuum model ansatz captures these additional details.

To represent the internal free energy of an ionic system comprising several species
of charge $Z_i$ and bulk concentrations $N_i$ each, we employ the exact expression
for the ideal gas of point particle ions, approximating finite-size effects with
a local density approximation, and write
\begin{equation}
\fl A_\kappa[\{\eta_i(\vec{r})\}] = T \sum_i \int\rmd\vec{r} N_i s(\vec{r})
\bigg[
	\underbrace{(\eta_i (\log \eta_i -1) + 1)}_{\textrm{\small Ideal gas}}
	+ \underbrace{\frac{(x(\vec{r})-x_0)^2}{x_0(1-x_0)^2(1-x(\vec{r}))^2}}_{\textrm{\small Hard sphere}}
\bigg].
\label{eqn:Akappa}
\end{equation}
The density of each ionic species is $N_i s(\vec{r}) \eta_i(\vec{r})$,
represented in terms of the enhancement $\eta_i(\vec{r})$ relative to
the cavity prescription of $N_i s(\vec{r})$. The charge-weighted sum of these densities contribute a net
ionic bound charge $\rho_\kappa = \sum_i Z_i N_i s(\vec{r}) \eta_i(\vec{r})$.

The first ideal gas term in (\ref{eqn:Akappa}) along with the mean-field electrostatic interaction
in the third term of (\ref{eqn:JDFT-PCM}) correspond to the Poisson-Boltzmann theory.
That theory, however, does not limit the density of the ions in solution and presents
an unphysical instability associated with infinite build-up of ions at regions of strong
external potential. We resolve this instability by enforcing a packing limit on the ions
via the second term of (\ref{eqn:Akappa}). This term captures local hard sphere correlations
in terms of the packing fraction $x(\vec{r}) = \sum_i V_i N_i \eta_i(\vec{r})$,
where $V_i = 4\pi R_i^3/3$ is the volume per ion for each species (with ionic radius $R_i$).
The functional form of this term is constrained to reproduce $x(\vec{r}) \to
x_0 \equiv \sum_i V_i N_i$ in the bulk and to match the divergence in the
equation of state of the hard sphere fluid \cite{CarnahanStarlingEOS} as $x \to 1$.

\subsection{Linear limit} \label{sec:LinearPCM}

The free energy functional (\ref{eqn:JDFT-PCM}) with dielectric free energy $A_\epsilon$
given by (\ref{eqn:Aeps}) and optional ionic free energy $A_\kappa$ given by (\ref{eqn:Akappa})
constitutes our nonlinear polarizable continuum model. Here, we show that the
conventional linear polarizable continuum model is a limit of this more general theory.

The rotational response from the dielectric is approximately linear
when the energy scale of the molecular dipoles interacting with the field
is much lower than the temperature ($p\sub{mol}|\vec{\nabla}\phi|\ll T$),
where $\phi(\vec{r})$ is the total electrostatic potential.
Similarly, the ionic response is approximately linear when $Z|\phi|\ll T$.
Using the Euler-Lagrange equations to eliminate $\vec{\varepsilon}(\vec{r})$ and $\mu(\vec{r})$
in favor of $\vec{\nabla}\phi(\vec{r})$ and $\phi(\vec{r})$ respectively,
expanding the free energy to quadratic order, and simplifying using
(\ref{eqn:epsBulk}) and the definition $\kappa^2 \equiv 8\pi N\sub{ion}Z^2/T$, we find
\begin{equation}
A_\epsilon + A_\kappa = \frac{1}{4\pi} \int \rmd\vec{r} s(\vec{r})
	\left[ (\epsilon_b-1)\frac{|\vec{\nabla}\phi|^2}{2} + \kappa^2 \frac{\phi^2}{2} \right],
\label{eqn:AdielLinear}
\end{equation}
with the corresponding total bound charge at linear order
\begin{equation}
\rho\sub{lq}(\vec{r})
	=  \frac{1}{4\pi}\left[ (\epsilon_b-1) \vec{\nabla} \cdot (s(\vec{r}) \vec{\nabla}\phi)
	- \kappa^2 s(\vec{r})\phi \right].
\label{eqn:rhoDielLinear}
\end{equation}

The Euler-Lagrange equation for this simplified linear-response
functional in terms of the single independent variable, $\phi$,
can be rearranged into the familiar modified Poisson equation
(or Helmholtz equation for non-zero $\kappa$)
\begin{equation}
\fl\qquad \vec{\nabla}^2 \phi(\vec{r})
+ (\epsilon_b-1)\vec{\nabla}\cdot (s(n(\vec{r})) \vec{\nabla}\phi(\vec{r}))
- \kappa^2 s(n(\vec{r})) \phi(\vec{r})
= -4\pi\rho\sub{el}(\vec{r}).
\label{eqn:PCM-Helmholtz}
\end{equation}
Finally, substituting the solution of (\ref{eqn:PCM-Helmholtz}) in
the fluid free energy functional (\ref{eqn:JDFT-PCM}) with the
dielectric and ionic energies given by (\ref{eqn:AdielLinear}),
yields an equilibrium value for $A\sub{diel}$ in the linear limit
\begin{equation}
A\sub{diel}\super{(linear)} = A\sub{cav}
 + \frac{1}{2} \int \rmd\vec{r} \rho\sub{el}(\vec{r})
	\left( \phi(\vec{r}) - \int \rmd\vec{r}' \frac{\rho\sub{el}(\vec{r}')}{|\vec{r}-\vec{r}'|} \right).
\label{eqn:AdielLinearPCM}
\end{equation}
Thus, the free energy functional approach to polarizable continuum models
reduces, in the linear limit, to the standard approach \cite{PCM-Kendra,PCM-Marzari}
of replacing the vacuum Poisson equation with one modified by the fluid.

\subsection{Periodic systems and net charge} \label{sec:ChargeNeutralityPW}

An important class of applications of the nonlinear polarizable continuum model,
and joint density-functional theory in general, is the study of electrochemical systems.
These systems pose an interesting challenge as they often involve charged metal
or doped semiconducting surfaces. The periodic boundary conditions necessary to
accurately describe the delocalized electronic states of such systems
complicate the addition of charge, since the energy per unit cell
of a periodic system with net charge per unit cell is divergent.

Including a counter electrode \cite{CounterElectrode} to keep the simulation cell neutral 
avoids this problem, but leads to wasted computational effort on irrelevant portions
of the system and complicates the separation of physics at the two electrodes.
Introducing Debye screening due to ions in the electrolyte neutralizes the
unit cell with fluid bound charge and naturally captures the physics of the
electrochemical double layer \cite{PCM-Kendra}. More importantly, unlike the
Poisson equation obtained without ionic screening, the Helmholtz equation
for the electrostatic potential with screening (\ref{eqn:PCM-Helmholtz}) has
a well-defined constant offset (`zero' of potential) in periodic boundary conditions.
The resulting Kohn-Sham eigenvalues, and hence the electron chemical potential,
correspond to a zero reference deep within the fluid, and this enables calibration
of the electron chemical potential in DFT against electrochemical reference electrodes.
(See \cite{PCM-Kendra} for details.)

The electrostatic potential in the nonlinear polarizable continuum model
is not obtained from a Helmholtz equation, and the bound charge in the ionic system
does not neutralize a net charge in the electronic system at an arbitrary
value of the independent variable $\mu(\vec{r})$. Here, we present the modifications
required to correctly handle periodic systems within the nonlinear ionic screening model.

The mean-field Coulomb energy per unit cell of volume $\Omega$ for the entire system
with total charge density $\rho\sub{tot} = \rho\sub{el} + \rho\sub{lq}$
can be written in the plane-wave basis as $U = \frac{1}{2\Omega}
\sum_{\vec{G}} \tilde{K}_{\vec{G}} |\tilde{\rho}\sub{tot}(\vec{G})|^2$.
Here, $\tilde{\rho}\sub{tot}(\vec{G}) = \int_\Omega \rmd\vec{r} \rho\sub{tot}(\vec{r})
\exp(-i\vec{G}\cdot\vec{r})$ for reciprocal lattice vectors $\vec{G}$,
and $\tilde{K}_{\vec{G}} = 4\pi/G^2$ is the plane-wave basis Coulomb kernel.
The divergent contribution at $G=0$ vanishes for neutral unit cells with
$Q\sub{tot}\equiv\tilde{\rho}\sub{tot}(0) = 0$.

The $G^{-2}$ divergence results from the long-range $1/r$ tail of the Coulomb kernel.
We can analyze the effect of the divergence by making the Coulomb kernel short-ranged
on a length scale $L$ much larger than the unit cell, and set $L\to\infty$ at the end.
The exact form of the regularization is not important; picking the Gaussian-screened
potential $\textrm{erfc}(r/L)/r$ results in the regularized Coulomb energy
\begin{equation}
U_L = \sum_{\vec{G}\neq 0} \frac{2\pi}{G^2\Omega} |\tilde{\rho}\sub{tot}(\vec{G})|^2
	+ \frac{\pi L^2}{2\Omega} Q\sub{tot}^2.
\label{eqn:U_L}
\end{equation}
Note that the first term employs the standard Coulomb kernel in the plane-wave basis
which drops the $G=0$ term by invoking a neutralizing background, and the second term
contains the divergent part depending only on the total charge per unit cell.

At finite large $L$, the second term of $U_L$ penalizes $Q\sub{tot} \neq 0$ and
favors equilibrium configurations with small $Q\sub{tot}$. The Euler-Lagrange
equation for the net charge $Q\sub{tot}$ is $\lambda \equiv \partial A/\partial Q\sub{tot}
= -\pi L^2 Q\sub{tot}/\Omega$, where $A$ is the total free energy excluding the divergent
second term of (\ref{eqn:U_L}). Note that $\partial A/\partial Q\sub{tot}$ is finite
for systems capable of adjusting their total charge, such as fluids with ionic screening, so that
as $L\to\infty$, $Q\sub{tot}\to 0$ in such a manner that $\lambda \propto Q\sub{tot}L^2$ remains finite.
The absolute potential is also well defined in this situation with a $G=0$ contribution
of $\partial U_L/\partial Q\sub{tot} = \pi L^2 Q\sub{tot}/\Omega = -\lambda$.
Finally, note that
\begin{equation}
U_\infty = \sum_{\vec{G}\neq 0} \frac{2\pi}{G^2\Omega} |\tilde{\rho}\sub{tot}(\vec{G})|^2
	-\lambda Q\sub{tot}
\end{equation}
results in the same Euler-Lagrange equation and equilibrium free energy as (\ref{eqn:U_L})
in the $L\to\infty$ limit, and therefore the divergent term in the Coulomb energy
reduces to a charge-neutrality constraint imposed by Lagrange multiplier $\lambda$.

We incorporate this Lagrange multiplier constraint into the ionic free energy in plane-wave calculations,
and retain the standard plane-wave Coulomb kernel with $G=0$ projected out for all electrostatic interactions.
The constraint can be solved analytically for local nonlinear ions (section~\ref{sec:NonlinearIons})
in the commonly encountered case of a `$Z$:$Z$' electrolyte consisting of two species of charge $+Z$ and $-Z$
(labeled with indices $i=+,-$) with bulk concentrations $N\sub{ion}$ each.
In this situation, we can show that substituting $\eta_{\pm}(\vec{r}) = \exp(\pm(\mu_0+\mu_\pm(\vec{r})))$,
where $\mu_0 \equiv -Z\lambda/T$ is obtained by solving the neutrality constraint, reduces the constrained
minimization over $\eta_{\pm}(\vec{r})$ to an unconstrained minimization over $\mu_\pm(\vec{r})$. 
In particular, the neutrality constraint $Q_+ e^{\mu_0} + Q_- e^{-\mu_0} + Q\sub{el} = 0$ yields
\begin{equation}
\mu_0 = \log\frac{\sqrt{Q\sub{el}^2 - 4 Q_+ Q_-} - Q\sub{el}}{2Q_+},
\end{equation}
where $Q_\pm \equiv \pm N\sub{ion}Z\int \rmd\vec{r} s(\vec{r}) e^{\pm \mu(\vec{r})}$ and
$Q\sub{el} \equiv \int \rmd\vec{r}\rho\sub{el}(\vec{r})$ is the total charge of the electronic system.
In this case, and for other joint density-functional theories which include ionic screening,
the constraint contribution to $\delta A\sub{diel}/\delta\rho\sub{el}(\vec{r})$ in the
electron potential establishes the absolute reference for the Kohn-Sham eigenvalues and the
electron chemical potential required for \emph{ab initio} electrochemistry \cite{PCM-Kendra}.

\subsection{Implementation} \label{sec:Implementation}

The nonlinear polarizable continuum model presented here and
its linear counterpart have been implemented in the open source
plane-wave electronic structure software JDFTx \cite{JDFTx},
designed for joint density-functional theory.
The electronic density-functional theory segment of this software
is based on conjugate gradients minimization \cite{PolakRibiereCG}
of an analytically continued total energy functional \cite{DFT-CG},
expressed in the DFT++ algebraic formulation \cite{DFT++}.
The fluid segment of JDFTx also employs the plane-wave basis and is discretized in the
algebraic formulation for classical density-functional theories \cite{RigidMoleculeCDFT}.

The valence electron density $n(r)$ from standard pseudopotentials need to be
augmented with a core electron density to prevent overlap of the fluid
with the pseudopotential cores \cite{PCM-Kendra}. Hence, we compute the
shape function using (\ref{eqn:ShapeFunction}) with $n\sub{cav}(\vec{r})
= n(\vec{r}) + n\sub{core}(\vec{r})$, where $n\sub{core}$ is the
partial core density used for nonlinear core corrections \cite{pcc}.

The electrostatic interactions with the fluid involve the total charge density
(both electronic and nuclear) of the material described in the electronic structure
portion of the calculation, $\rho\sub{el}(\vec{r}) = n(\vec{r}) + \rho\sub{nuc}(\vec{r})$.
Here, the nuclear charge density, $\rho\sub{nuc}(\vec{r}) = -
\sum_i Z_i e^{-(\vec{r}-\vec{r}_i)^2/(2w^2)} / (2\pi w)^{3/2}$ \footnote{
Note that we employ an electron-is-positive charge convention, so that
$\rho\sub{nuc} < 0$ and the charge of the electron is +1 in atomic units.}
is widened by a Gaussian resolvable on the charge density grid.
The widened nuclear density is used only in the interaction with the fluid;
the internal energies of the electronic system employ point nuclei in all terms.
This width does not affect the interaction energy since the fluid and nuclear
charge densities do not overlap, and the nuclear charge is spherically symmetric.
However, it shifts the potential relative to the zero-width case, which we
compensate exactly by adding the correction $-2\pi w^2 \sum_i Z_i/\Omega$
to the electron potential, where $\Omega$ is the unit cell volume.

Finally, regarding algorithms, the linear polarizable continuum models are minimized
by solving the Helmholtz (or Poisson) equation (\ref{eqn:PCM-Helmholtz}) at every
electronic iteration. Appropriate preconditioners for the involved
linear conjugate gradients solver have been developed previously \cite{PCM-Kendra}.
The free energy of the nonlinear polarizable continuum model is
minimized using the Gummel iteration \cite{GummelLoop}, where
the electronic system and the fluid are alternately minimized
while holding the state of the other one frozen. This method is
guaranteed to be globally convergent due to the variational principle,
and typically converges adequately in 5-10 alternations for most systems studied.
The fluid free energy $A\sub{diel}$ is minimized with the scalar field
$\mu(\vec{r})$ and vector field $\vec{\varepsilon}(\vec{r})$ as independent variables;
the diagonal preconditioner in reciprocal space \footnote{This preconditioner
is derived from an approximation to the Hessian of $A\sub{diel}$ with respect
to $\mu(\vec{r})$ and $\vec{\varepsilon}(\vec{r})$ in the bulk linear limit.}
\begin{equation}
K_\mu(\vec{G}) = \left[\frac{Z(1-\alpha/3)}{p\sub{mol}}\right]^2
	\frac{G^2}{(G^2 + \kappa^2/\epsilon_b)^2}
\end{equation}
for the $\mu$ channel with the identity preconditioner on the
$\vec{\varepsilon}$ channel yields satisfactory convergence
for the nonlinear conjugate gradients algorithm \cite{PolakRibiereCG}.

\section{Results}

Strong electric fields at liquid interfaces typical of battery systems
necessitate a theory for the nonlinear response of the liquid environment,
such as the nonlinear polarizable continuum model of section~\ref{sec:Theory}.
Section~\ref{sec:Molecules} calibrates the undetermined parameters of
this theory against experimental solvation energies of molecules.
For these molecules, and for metallic surfaces in section~\ref{sec:MetalSurfaces},
we find results comparable to linear PCM's. However, for surfaces of ionic solids
in section~\ref{sec:IonicSurfaces}, we find that inclusion of nonlinear effects
are necessary in order to obtain qualitatively correct results.

\subsection{Computational Details}

We perform all calculations in this paper using the open source plane-wave
density functional software JDFTx \cite{JDFTx} at a plane wave cutoff
of 30 $E_h$ (1 $E_h \equiv$ 1 hartree $\approx$ 27.21 eV).
These calculations employ norm-conserving pseudopotentials generated by the Opium pseudopotential
generator \cite{opium} with the PBE exchange and correlation functional \cite{PBE}.
The pseudopotentials for metal atoms include partial core corrections \cite{pcc},
which are necessary to keep the fluid out of the pseudopotential cores
as described in Section~\ref{sec:Implementation}.

The choice of exchange-correlation functional for molecular and surface systems
is not straightforward \cite{DFT_SurfaceConsiderations}, and some argue that
semi-local approximations can be inadequate for these systems \cite{QMCsurfaceThesis}.
Hybrid functionals which include exact exchange, or quantum Monte Carlo methods,
are likely to be more accurate but are significantly more expensive than
semi-local methods and hence unsuitable for rapid screening calculations.
Here, we use the semi-local revTPSS exchange-correlation functional \cite{revTPSS}
which shows considerable promise for accurate calculations of surface phenomena
including surface formation energies and molecular adsorption energies \cite{DFTbenchmarkSurfaces}.

Molecular geometries for the calculations of section~\ref{sec:Molecules}
are from the Computational Chemistry Comparison and Benchmark Database \cite{cccbdb}.
The surface geometries employed in sections~\ref{sec:IonicSurfaces} and \ref{sec:MetalSurfaces}
are constrained to the optimized bulk geometry for the central layer, while
the remaining layers are fully relaxed for both the vacuum and fluid calculations.
The fluid models assume ambient temperature $T = 298$~K for all calculations.

\subsection{Calibration to molecular solvation energies} \label{sec:Molecules}

The nonlinear dielectric response of section~\ref{sec:NonlinearDielectric} is
completely constrained by \emph{ab initio} and experimentally determined parameters,
listed in table~\ref{tab:solvents} for the solvents studied in this paper.
However, the cavity shape function and the cavitation and dispersion terms,
which are integral features of any polarizable continuum model, are unknown
microscopic quantities that are typically constrained by a fit to solvation energies.
Here, we fit the set of unknown cavity parameters for the nonlinear model and its linear limit
to the same molecular solvation dataset using the same procedure, in order to facilitate
a fair comparison between linear polarizable continuum models and our nonlinear theory.

\begin{figure}
\center{\includegraphics[width=\columnwidth]{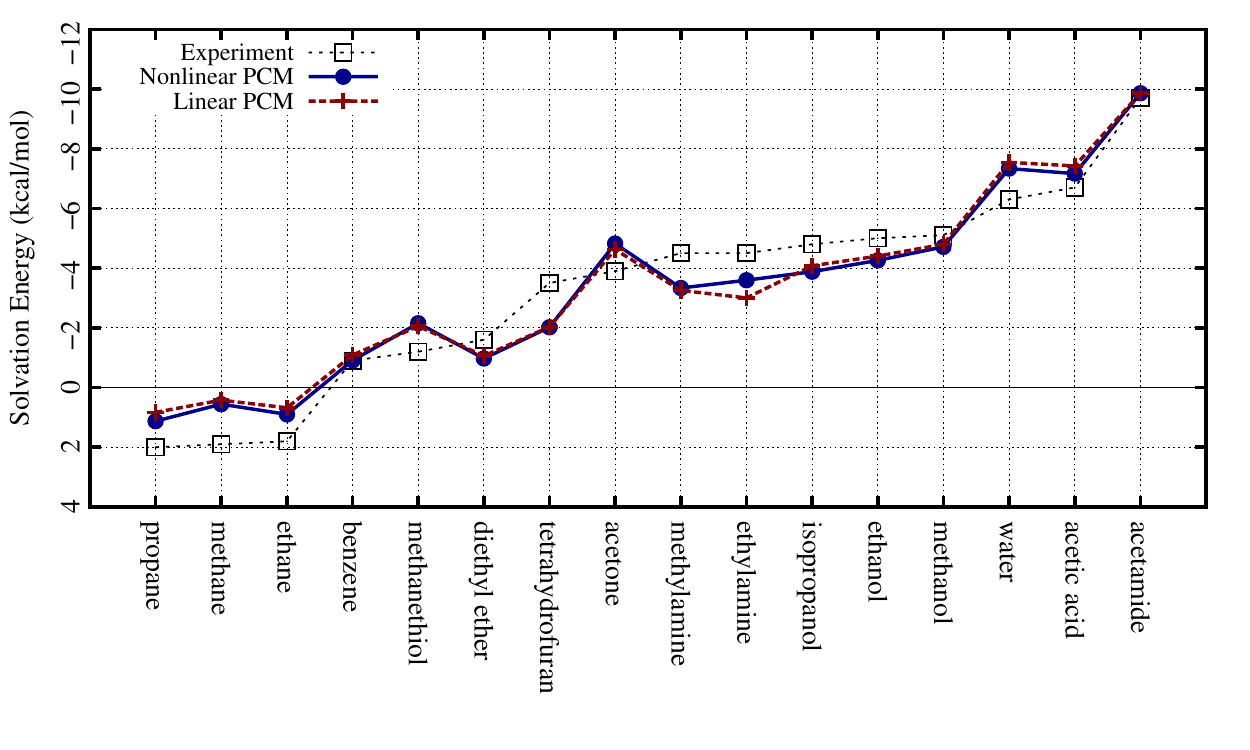}}
\caption{Solvation energies of molecules in water predicted by the
nonlinear and linear polarizable continuum models compared against the
experimental values from \cite{solvation-exp-compiled-1,solvation-exp-compiled-2}.
\label{fig:solvation_energies}}
\end{figure}

The molecular solvation dataset must contain experimental data that is both reliable and readily available.
Organic molecules solvated in water satisfy this criterion and are commonly used in fitting
parameters for polarizable continuum models \cite{PCM-Marzari,PCM-Review}.
The molecules used in our fit are listed in figure~\ref{fig:solvation_energies},
and the known solvent parameters for water are listed in table~\ref{tab:solvents}.
Of the remaining parameters, we set the shape function width
parameter $\sigma = 0.6$ as in \cite{JDFT,PCM-Kendra}
since the solvation energies are somewhat insensitive to it.
We then determine the cavity transition electron density $n_c$ and
the effective cavity tension $\tau$ by a nonlinear least squares
fit to the molecular solvation energy dataset.

\begin{table*}
\begin{center}
\begin{tabular}{lccc}
	\hline
	\hline
		& $n_c$ ($a_0^{-3}$) & $\tau$ ($E_h/a_0^2$) & RMS Error (kcal/mol)\\
	\hline
	Nonlinear PCM & \sci{1.0}{-3} & \sci{9.5}{-6} & 0.95 \\
	Linear PCM    & \sci{3.7}{-4} & \sci{5.4}{-6} & 1.05 \\
	\hline
	\hline
\end{tabular}
\end{center}
\caption{Fitted parameters of the nonlinear and linear polarizable continuum models (PCM)
and the corresponding RMS errors for solvation energies of the molecules listed in
figure~\ref{fig:solvation_energies}.
\label{tab:fit_results}}
\end{table*}

\begin{figure}
\includegraphics[width=0.96\columnwidth]{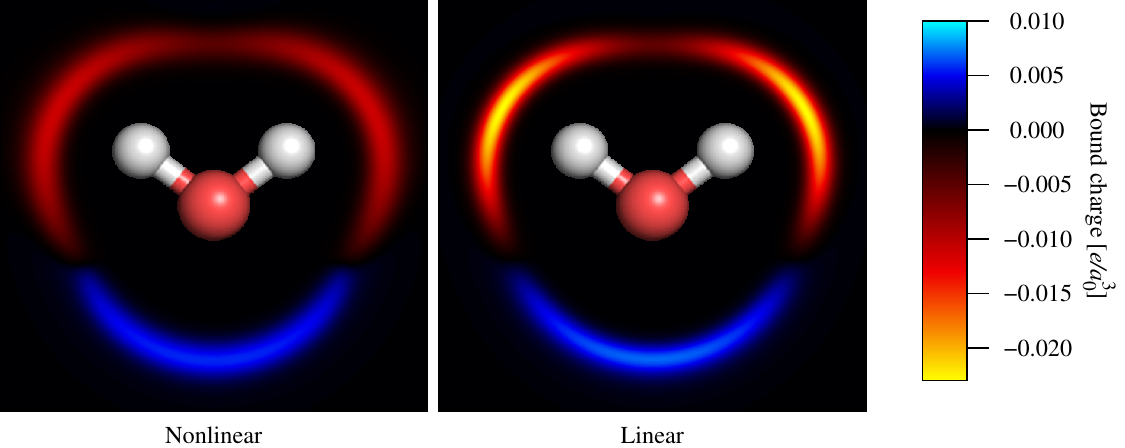}
\caption{Bound charge in solvent water around a water molecule in the nonlinear and linear models.
The smaller hydrogen atoms produce stronger fields on the solvent compared to the oxygen,
resulting in much stronger saturation effects in the negative bound charge surrounding the hydrogens.
In spite of the increased bound charge, the linear model yields approximately the same
solvation energy as the nonlinear one due to compensation by the increased cavity size.
\label{fig:waterBoundCharge}}
\end{figure}

The resulting fit parameters and optimized RMS error in solvation energy for the
nonlinear and linear versions of the model are summarized in table~\ref{tab:fit_results}.
The smaller $n_c$, and hence larger cavities, for the linear model as compared to the nonlinear one
offset the overestimation of electrostatic interactions due to the lack of saturation effects.
The lowered cavity tension $\tau$ in the linear model then compensates for the increase in cavity area.
Figure~\ref{fig:waterBoundCharge} demonstrates the consequences of these differences
in the solvent bound charge surrounding a water molecule. The solvation energies predicted by
the two models are in agreement as seen in figure~\ref{fig:solvation_energies},
in spite of significantly larger bound charges in the linear case.
Due to this cancellation, the linear model yields comparable accuracy to the nonlinear one
for the solvation of organic molecules in water, but this is no longer the case
when stronger electric fields come into play, as in some of the electrochemical systems
we study next.

\subsection{Solvation of metallic surfaces} \label{sec:MetalSurfaces}

\begin{figure}
\includegraphics[width=0.44\columnwidth]{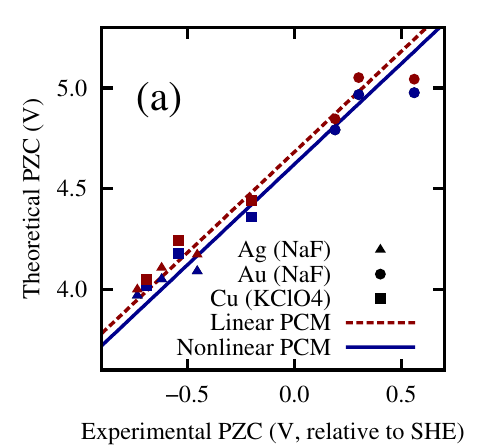}
\includegraphics[width=0.55\columnwidth]{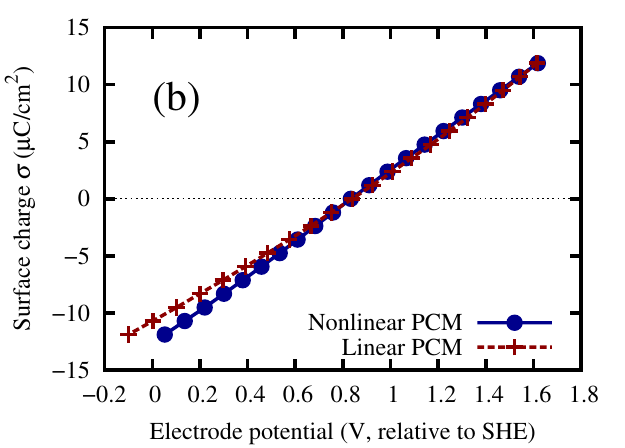}
\caption{
(a) Potentials of zero charge (PZC's) for the (111), (100) and (110) (left to right)
each for silver (circles), gold (triangles), and copper (squares), predicted by
the nonlinear and linear theories, compared to experiment \cite{PZC}.
The diagonal line for each theory compares theoretical and experimental values
up to an overall fitted offset. (See table~\ref{tab:SHEvalues}.)
The silver and gold are solvated in aqueous 1~M NaF electrolyte (ionic radii Na: 1.16~\AA~ F: 1.19~\AA),
while the copper is in aqueous 1~M KClO$_{4}$ electrolyte (ionic radii K: 1.52~\AA~ and ClO$_4$: 2.26~\AA).
(b) Charge on a Pt(111) surface in 1~M aqueous KClO$_4$ as a function of
potential relative to the standard hydrogen electrode (SHE) for the two theories.
\label{fig:charge_vs_V}
\label{fig:PZC}
}
\end{figure}

Unlike the typical electrochemical interface, noble metal electrodes in
electrolyte are less prone to complex chemical interactions at the surface,
making them suitable candidates for an initial evaluation of our theory.
Reactions are highly sensitive to the absolute electron chemical potential,
which in experiments is typically reported relative to the
standard hydrogen electrode (SHE). The absolute potential of the SHE
relative to vacuum is difficult to establish experimentally;
the estimates from different experimental methods range from 4.4~V to 4.9~V \cite{PZC}.
To make direct contact with experimental electrochemical observables,
this experimentally uncertain quantity can be calibrated \cite{PCM-Kendra} in
density-functional theory by comparing the theoretical chemical potentials for solvated
neutral metal surfaces against the experimental potentials of zero charge (PZC's).
The calibrations of the reference electrode potential within the linear and nonlinear models
are remarkably similar, as shown in figure~\ref{fig:PZC}(a) and table~\ref{tab:SHEvalues}.

\begin{table*}
\begin{center}
\begin{tabular}{lccc}
	\hline
	\hline
		& $V$\sub{SHE} (V) & $V$\sub{dip} (V) & RMS Error (V)\\
	\hline
	Nonlinear PCM   & 4.62 & 0.46 & 0.09 \\
	Linear PCM    & 4.68 & 0.40 & 0.09 \\
	\hline
	\hline
\end{tabular}
\end{center}
\caption{Offset between theoretical and experimental PZC's, $V$\sub{SHE},
determined by a fit using the systems of figure~\ref{fig:PZC}(a), with corresponding RMS errors.
$V$\sub{SHE} represents the potential difference between an electron solvated deep in the fluid
and the Standard Hydrogen Electrode. $V$\sub{dip} represents the potential due to the dipole moment
at the fluid-metal interface, and is obtained as the difference between the theoretical PZC
and the work function, averaged over the systems considered for each fluid model.
\label{tab:SHEvalues}}
\end{table*}

The absolute potential of zero charge includes contributions from
the work function, which is essentially independent of the fluid theory,
and from the dipole moment in the interfacial layers of the liquid.
The minor differences in the calibrations of the two theories
stem from this dipole moment contribution, as shown for
aqueous electrolytes in table~\ref{tab:SHEvalues}.
The variation of surface charge with electrode potential is
also similar for the two models, as shown for the solvated Pt(111)
surface of figure~\ref{fig:charge_vs_V}(b).
In particular, the derivative of that variation, the so-called
`double-layer' capacitance, at the potential of zero charge
is 14 and 15~$\mu$F/cm$^2$ for linear and nonlinear PCM respectively,
which agrees well with an experimental estimate of 20~$\mu$F/cm$^2$
\cite{DoubleLayerCapacitance} for the above system.

The agreement in the results of the linear and nonlinear theories
demonstrated in figures~\ref{fig:charge_vs_V}(a,b) and table~\ref{tab:SHEvalues}
is due to the same cancellation of errors at play for solvation of molecules.
The linear theory misses saturation in the rotational dielectric response,
thereby overestimating it, yet compensates with an increase in cavity size.
This cancellation of errors is possible since the typical magnitudes
of electric fields under typical operating potentials are similar
to those of the molecular case, as shown in figure~\ref{fig:polfrac}.

Both models predict an approximately linear variation of surface
charge with electrode potential (figure~\ref{fig:charge_vs_V}(b)),
which corresponds to a constant capacitance.
This prediction contrasts with the experimental observation of a capacitance
minimum at the potential of zero charge \cite{DoubleLayerCapacitance}
due to ion adsorption on the electrode surface.
The formation of this so-called inner Helmholtz layer between the solid surface and
the solvent is precluded by the cavity ansatz of polarizable continuum models.
These details require either a higher level of theory capable of describing
layering effects of ions such as a classical density-functional approach,
or the inclusion of explicit ions into the quantum mechanical calculation.
Nonetheless, both the linear and nonlinear PCM adequately describe
the basic features of the ideal electrochemical interface, and 
are suitable for describing chemical reactions at metal electrode surfaces
as long as all chemical bonds are treated quantum-mechanically.

\subsection{Solvation of ionic surfaces} \label{sec:IonicSurfaces}

The surfaces of electrodes typically contain ionic compounds whose structure
and composition vary with the chosen electrolyte and operating conditions.
The details of this interface play a critical role in battery performance,
and an accurate description of such surfaces in electrolyte environments
is therefore crucial for modeling efforts towards improving battery systems.
Reactions at the surface of a lithium metal anode, for example, can form
Li\sub{2}O, LiOH and LiF at the solid electrolyte interface \cite{HF_SEI,SEI_general}.
Here, we study these surfaces in contact with different organic solvents typical
of battery systems as a testbed for fluid models applicable to battery systems.

\begin{table*}
\begin{center}
\begin{tabular}{ccccccc}
    \hline
    \hline
    Solvent & $\epsilon_b$ & $\epsilon_\infty$ & $p\sub{vac}$ ($e a_0$) & $p\sub{mol}$ ($e a_0$) & $N\sub{mol}$ ($a_0^{-3}$) & $\tau$ ($E_h/a_0^2$) \\
    \hline
    Water & 78.4 & 1.78 & 0.73 & 0.94 & \sci{4.938}{-3} & \sci{9.5}{-6} \\
    \hline
    DMC   &  3.1 & 1.87 & 0.16 & 0.16 & \sci{1.059}{-3} & \sci{2.05}{-5} \\
    THF   &  7.6 & 1.98 & 0.69 & 0.90 & \sci{1.100}{-3} & \sci{1.78}{-5} \\
    DMF   & 38.0 & 2.05 & 1.50 & 2.19 & \sci{1.153}{-3} & \sci{2.26}{-5} \\
    PC    & 64.0 & 2.02 & 1.97 & 2.95 & \sci{1.039}{-3} & \sci{2.88}{-5} \\
    EC    & 90.5 & 2.00 & 1.93 & 2.88 & \sci{1.339}{-3} & \sci{3.51}{-5} \\
    \hline
    \hline
\end{tabular}
\end{center}
\caption{Parameters describing water and commonly used lithium battery solvents, Dimethyl Carbonate (DMC),
Tetrahydrofuran (THF),  Dimethylformamide (DMF), Propylene Carbonate (PC) and Ethylene Carbonate (EC).
The vacuum dipoles ($p\sub{vac}$) and self-consistent solvated dipoles $p\sub{mol}$ are computed
using density-functional theory as described in section \ref{sec:NonlinearDielectric}.
All remaining parameters are constrained by measured bulk properties \cite{springer-materials}.
\label{tab:solvents}}
\end{table*}

The solvents selected for this study are listed in table~\ref{tab:solvents}.
Due to the dearth of experimental data for corresponding solvation energies, 
we  here use the cavity shape parameters determined by the fit in section~\ref{sec:Molecules}.
We replace the effective tension $\tau$ by the experimental surface tension,
ignoring dispersion effects which are insignificant on the scale of the electrostatic
energies in these highly polar systems. All remaining physical parameters that  
determine the dielectric response are constrained by experiment and \emph{ab initio}
calculations, as discussed in section~\ref{sec:NonlinearDielectric}.

\begin{figure}
\center{\includegraphics[width=\columnwidth]{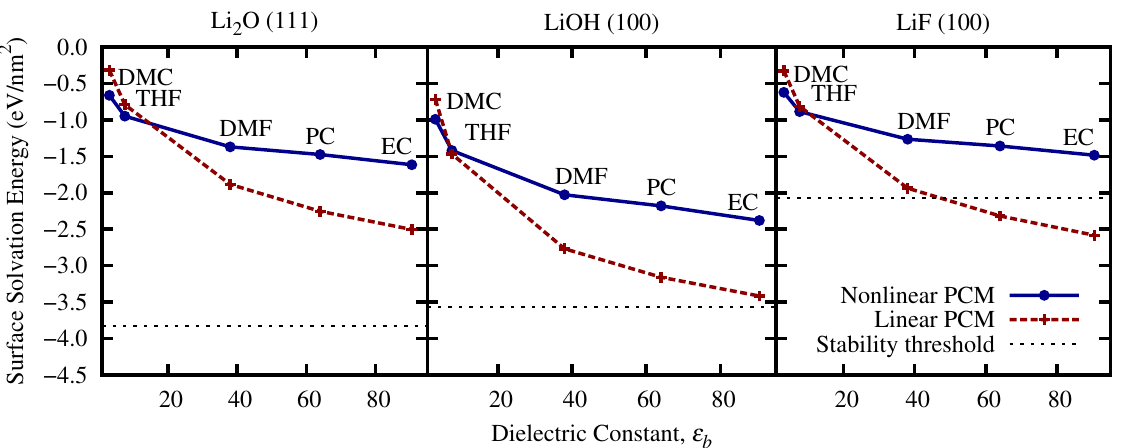}}
\caption{Solvation energies predicted by the nonlinear and linear models for
surfaces of Li\sub{2}O, LiOH and LiF in the organic solvents from table~\ref{tab:solvents}.
With increasing dielectric constant, the predictions of the linear model
diverge from those of the nonlinear model due to missing saturation effects.
This leads to qualitative differences here, unlike the case
of the solvated molecules of figure~\ref{fig:solvation_energies}.
For some surfaces, the linear model suggests, perhaps incorrectly, that the flat surface
is unstable by lowering the solvated surface energy relative to the bulk solid.
\label{fig:SurfaceEnergies}}
\end{figure}

\begin{figure}
\center{\includegraphics[width=0.84\columnwidth]{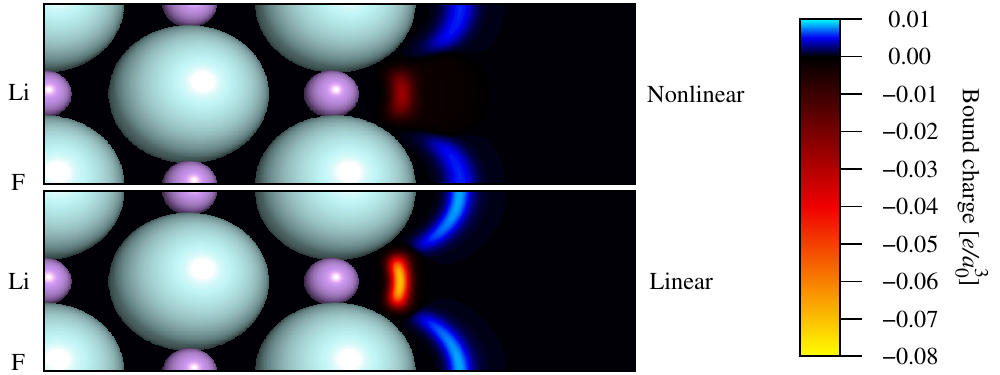}}
\caption{Bound charge in solvent Ethylene Carbonate (EC) around a LiF (100) surface
in the nonlinear and linear models, shown in a (011) slice. Saturation effects
are stronger next to the smaller Li$^+$ cations which produce significantly
stronger fields on the solvent compared to the larger F$^-$ anions.
In contrast to a water molecule solvated in water (Figure~\ref{fig:waterBoundCharge}),
these effects are strong enough to qualitatively alter
solvation energies, as shown in figure~\ref{fig:SurfaceEnergies}.
\label{fig:LiFsurfaceBoundCharge}}
\end{figure}

The linear and nonlinear models predict similar solvation energies for the aforementioned
ionic compounds of lithium in solvents with low dielectric constants, as shown in
figure~\ref{fig:SurfaceEnergies}. However, with increasing dielectric constant,
the magnitude of the solvation energy increases more rapidly for the linear model,
leading to disagreement by up to a factor of two for the most polar solvents.
The linear model overestimates the electrostatic interaction due to a lack of saturation effects,
but unlike in the molecular case, the increase in cavity size is insufficient to compensate for this error.
In fact, for lithium fluoride in ethylene carbonate, as seen in figure~\ref{fig:LiFsurfaceBoundCharge}, 
the linear model model overestimates the bound charge by an order of magnitude.  
Indeed, in this case, the result is a qualitative difference in the predicted stability of the solvated surface
relative to the solid, with the linear model even predicting the solid to be 
thermodynamically unstable with respect to the formation of surfaces in this system.

\begin{figure}
\center{\includegraphics[width=0.8\columnwidth]{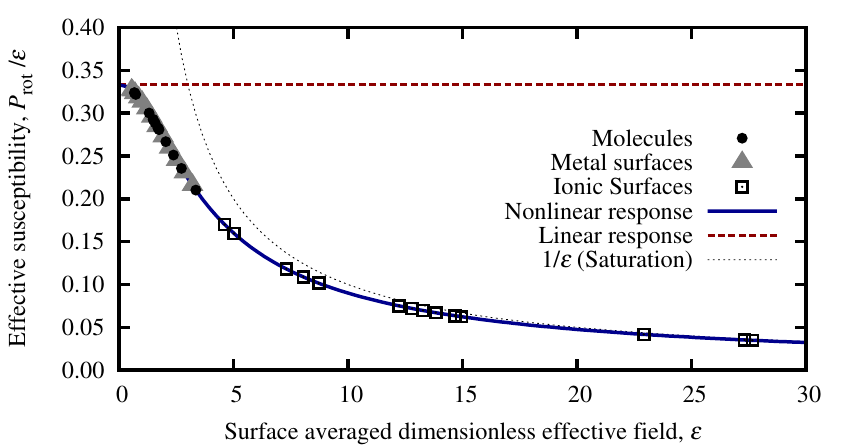}}
\caption{Effective rotational susceptibility at the average value of the dimensionless
effective field $\varepsilon$ at the cavity surface (solvent-solute interface) for
solvated molecules (circles), charged metal surfaces (triangles) and ionic surfaces (squares).
The reduction in susceptibility due to saturation effects captured by the nonlinear model
is missed by the linear one. Unlike the case of molecules and metal surfaces,
the order of magnitude overestimation of the susceptibility by the linear model
for ionic surfaces is not compensated by the increase in cavity size.
\label{fig:polfrac}}
\end{figure}

The qualitative inadequacy of the linear model for ionic surfaces derives from the significantly
stronger electric fields in these systems compared to solvated molecules and metallic surfaces.
Figure~\ref{fig:polfrac} compares the average electric field and the corresponding
rotational susceptibility at the solute-solvent interface for all the systems discussed above.
The least polar neutral ionic surface still imposes a higher electric field than the
most polar molecule or charged metallic surface at chemically relevant electrode potentials.
The order of magnitude reduction in rotational susceptibility due to saturation effects
in the ionic surfaces, compared to the modest reduction for the other systems,
necessitates a nonlinear theory for the study of these types of battery systems.

\section{Conclusions}

\emph{Ab initio} studies provide key insights into chemical processes
in a wide range of systems, but have not yet approached battery chemistry
with a realistic description of the electrolyte environment.
Continuum solvation models provide an intuitive and computationally-efficient
description of the environment and enable a focused study of the complex
subsystems that require treatment at the electronic structure level.
Our results indicate that standard polarizable continuum models fit
to molecular solvation data perform poorly when applied to polar
surfaces of the type often encountered at the SEI in battery systems.
Consequently, one must exercise caution when attempting to apply standard solvation
models available in both quantum chemical \cite{PCM-Review,Gaussian,GAMESS} and 
condensed matter \cite{PCM-Marzari,PWSCF} \emph{ab initio} software packages.
As an alternative, the nonlinear theory presented here and implemented in \cite{JDFTx}
leverages the computational simplicity of the standard polarizable continuum models
and extends their applicability to systems with the strong electric fields
associated with ionic surfaces in electrochemical systems.

The importance of nonlinear solvent response depends on the strength of electric
fields at the interface, which in turn varies dramatically with system type,
as highlighted in figure~\ref{fig:polfrac}. For systems with moderate
field strengths, such as the molecules and metal surfaces studied here,
the linear models can compensate for the overestimated electrostatic response
through an increase in cavity size. However, for systems with higher 
field strengths, such as ionic surfaces, this compensation is insufficient.
The nonlinear polarizable continuum model developed here consistently describes all of these systems,
and along with the technique developed in section~\ref{sec:ChargeNeutralityPW}
to determine the absolute electron chemical potential, enables electronic structure
predictions for real electrochemical systems as a function of electrode potential.

This theory provides a cost-effective yet accurate method for calculating
properties of electrochemical systems of technological relevance,
such as high-throughput screening potential battery materials.
Cleaner surface experiments analogous to 
the solvation datasets available for molecules will help 
further refine the theory of solvation for these systems.
In our study of battery electrode materials, we showcase our
theory with electronic density functional calculations.
The method can easily be used with other electronic theories such as
coupled cluster or quantum Monte Carlo \cite{Katie-QMC},
enabling the study of non-equilibrium properties.
Using these techniques, we can include nonlinear solvation 
in transition state calculations important for understanding
processes in energy systems, such as catalysis in fuel cells
and ion diffusion on battery electrodes.

\section*{Acknowledgements}
This work was supported as a part of the Energy Materials Center at Cornell (EMC$^2$),
an Energy Frontier Research Center funded by the U.S. Department of Energy,
Office of Science, Office of Basic Energy Sciences under Award Number DE-SC0001086.
D.G. also acknowledges financial support by DOE DE-FG02-07ER46432.
K.A.S and K.L.-W. were financially supported by National Science Foundation
Graduate Research Fellowships.  

\section*{References}
\bibliographystyle{unsrt}

\end{document}